\documentclass[12pt]{iopart}
\usepackage{graphicx}

\newcommand{\avg}[1]{\langle #1 \rangle}
\newcommand{\vi}{\mathbf{v}_i}
\newcommand{\ri}{\mathbf{r}_i}
\newcommand{\Peq}{P_{\rm eq}}
\newcommand{\Dt}{\Delta t}
\newcommand{\Dx}{\Delta x}
\newcommand{\Dr}{\Delta r}
\newcommand{\stdfigsize}{3.375in}
\newcommand{\widefigsize}{5.5 in}
\newcommand{\tx}[1]{{\rm #1}}

\begin{document}

\title[Dynamic heterogeneity and non-Gaussian behavior]{Dynamic heterogeneity and non-Gaussian behavior in a model supercooled liquid}

\author{M. Scott Shell\dag, Pablo G. Debenedetti\dag and Frank H. Stillinger\ddag}
\address{\dag Department of Chemical Engineering, Princeton University, Princeton, NJ 08544, USA}
\address{\ddag Department of Chemistry, Princeton University, Princeton, NJ 08544, USA}
\ead{shell@princeton.edu; pdebene@princeton.edu; fhs@princeton.edu}
\date{June 21, 2005}

\begin{abstract}
We use a recently-derived reformulation of the diffusion constant [Stillinger F H and Debenedetti P G 2005 \emph{J. Phys. Chem. B} \textbf{109} 6604] to investigate heterogeneous dynamics and non-Gaussian diffusion in a binary Lennard-Jones mixture.  Our work focuses on the joint probability distribution of particles with velocity $v_0$ at time $t=0$ and eventual displacement $\Dx$ at time $t=\Dt$.  We show that this distribution attains a distinctive shape at the time of maximum non-Gaussian behavior in the supercooled liquid.  By performing a two-Gaussian fit of the displacement data, we obtain, in a non-arbitrary manner, two diffusive length scales inherent to the supercooled liquid and use them to identify spatially separated regions of mobile and immobile particles.
\end{abstract}

\submitto{\JPCM}

\maketitle

\section{Introduction}
\label{sec:intro}

Perhaps no other manifestation of diffusive phenomena has received as much scientific attention in recent years as the nature of molecular motions in supercooled liquids.  When cooled sufficiently below their freezing point (supercooled), liquids exhibit a wealth of seemingly anomalous features not observed in stable, ``hot'' liquids \cite{EN1524, EN932}.  In deeply supercooled liquids, the Stokes-Einstein dependence of the translational diffusion coefficient on viscosity breaks down, with molecules actually translating faster than predicted by this venerable equation \cite{EN3062, EN939}.  No such breakdown appears in the rotational diffusion coefficient.  Moreover, supercooled liquids develop two characteristic relaxation times \cite{EN3064, EN3069}: a fast $\beta$ relaxation corresponding to highly local atomic movements, and a much slower $\alpha$ relaxation identified with major configurational rearrangements.  For the former, it is found that the temperature dependence is Arrhenius and the decay of correlations is exponential, while the latter exhibits highly non-Arrhenius behavior and its time dependence is often better captured by a ``stretched'' exponential form.

One of the most intriguing aspects of deeply supercooled liquids is so-called dynamic heterogeneity \cite{EN939, EN1588, EN3151, EN3190}.  This term refers to the existence of spatially separated regions, typically on the order of several nanometers, whose relaxation dynamics can differ from each other by a factor of up to 5 orders of magnitude \cite{EN939}. Clearly, these regions are transient, since over very long periods of time each particle will perform a similar average trajectory.  Dynamic heterogeneity therefore supposes that when viewed over an intermediate time scale, particles can be classified into mobile and immobile groups which appear to cluster together.  For supercooled liquids only a few percent above their glass transition temperature, such correlations persist long enough to be observed experimentally.  Indeed, heterogeneous dynamics in supercooled liquids have been identified in numerous experiments, as well as in computer simulations (see, for example, the references within \cite{EN939, EN1588, EN3151}).

The fundamental question regarding dynamic heterogeneity is: what is its origin and, thus, to what degree is it universal?  Some established theories of supercooled liquid dynamics---such as the mode-coupling \cite{EN3155, EN3156} and free-volume \cite{EN3159} approaches---treat the liquid as homogeneous and therefore cannot comment on dynamic heterogeneity.  The entropy-based theory of Adam and Gibbs \cite{EN1593}, on the other hand, invokes the notion of a ``cooperatively rearranging region'' but provides no basis for its calculation.

Recent studies have attempted to understand dynamic heterogeneity, both theoretically and using molecular dynamics simulations, although a definitive explanation has yet to emerge.  The theory of ``dynamic facilitation,'' for instance, envisions mobile particles in the liquid that can excite neighboring particles and make them mobile as well \cite{EN2917, EN2916}.  This theory has received support from simulation \cite{EN3091} and has offered a fruitful perspective for understanding the consequences of dynamic heterogeneity.  However, its presupposition of a particular relaxation mechanism does not provide insight into the molecular origins of dynamic heterogeneity.  Although other theories exist, a comprehensive survey of the current interpretations of dynamic heterogeneity is well beyond the scope of this paper.  We refer the interested reader to the reviews of Ediger \cite{EN939}, Glotzer \cite{EN1588}, and Richert \cite{EN3151}.

Computer simulations have been extremely useful for understanding dynamic heterogeneity, particularly in identifying mechanisms of cooperative rearrangement among mobile regions in the liquid \cite{EN3119, EN3118, EN3178, EN3189, EN3170, EN3192, EN3098, EN3075, EN3191}.  The primary focus of these studies has been the single-particle displacement vector, $\Delta \ri(\Dt)$, and its associated probability distribution, $P(\Delta \ri)$.  In the seminal work of Kob \etal \cite{EN3119}, it was found that dynamic heterogeneity in computer simulation of a binary glass-former was closely related to a non-Gaussian distribution of particle displacements.  These authors identified a time $\Dt^*$ at which $P(\Delta \ri)$ exhibited maximum deviations from a Gaussian distribution.  They defined a critical value of $\Delta r$ as the displacement for which the difference between the self part of the van Hove correlation function and the corresponding Gaussian approximation ``starts to become positive and very large.''  Kob \etal identified mobile particles as those whose displacement exceeded this distance during the interval $\Dt^*$, and they subsequently showed that this subset of particles was strongly correlated in space.  Similar studies have followed this initial effort.  One of the challenges that remains in such investigations (including ours) is the rigorous selection of an appropriate time interval for the calculation of displacements and of a method for classifying mobility based on $\Delta \ri$.  Ideally, one would like to make these choices based on physical arguments and the emergent properties of supercooled liquids alone, with no arbitrarily-determined parameters.

The goal of the present work is to investigate the extent to which a recently derived diffusion formalism \cite{EN3193} offers new insights into dynamic heterogeneity and non-Gaussian diffusion.  This formalism involves the calculation of the joint probability of initial particle velocity and final displacement after a given time interval, the long-time limit of whose cross-moment yields the diffusion coefficient.  Relative to previous studies, our approach considers the possible effects of a particle's initial velocity on heterogeneous diffusion.  We perform molecular dynamics calculations for the well-studied binary Lennard-Jones mixture \cite{EN3070} in the supercooled state.  Our results show that the joint distribution attains a qualitatively distinct form at the time of maximum non-Gaussian behavior, and this form in turn suggests a two-Gaussian parametrization of particle displacements.  By performing a two-Gaussian fit, we extract length scales which we use to identify spatially correlated regions of mobility and immobility in the liquid.  These findings offer an alternative but complementary point of view to current simulation studies of dynamic heterogeneity.

This paper is organized as follows.  Section \ref{sec:background} reviews the diffusion formalism proposed in Ref.~\cite{EN3193} and the analytical procedures for characterizing heterogeneous dynamics.  Section \ref{sec:results} presents the main results from our computer simulations, with Section \ref{sec:conclusion} commenting on their relevance and relationship to other ongoing work.

\section{Background and formalism}
\label{sec:background}

The central focus of Einstein's familiar expression for the self-diffusion constant $D$ is the rate of increase of the average squared particle displacement \cite{EN3194}, given by
\begin{equation}
\label{eq:diffmsd}
D = \lim_{t\to \infty} \frac{1}{6}\frac{d}{dt} \avg{[\Delta \ri(t)]^2}
\end{equation}
where $\ri(t)$ is the position of particle $i$ at time $t$ and $\Delta \ri(t) = \ri(t)-\ri(0)$.  The average is performed over all particles of a particular species and over all trajectories consistent with the thermodynamic state of the system (i.e., thermal averaging). Alternatively, one may move the derivative inside the average and use a change of time origin to obtain the equivalent velocity autocorrelation expression for the diffusion constant:
\begin{equation}
\label{eq:diffvac}
D = \frac{1}{3} \int_0^\infty \avg{\vi(0) \cdot \vi(t)} dt
\end{equation}
where $\vi(t)$ is particle $i$'s velocity.  Equations \ref{eq:diffmsd} and \ref{eq:diffvac} are well-established statistical mechanical formulas and are routinely used to determine self-diffusion constants in molecular dynamics simulations \cite{EN1598}.  It is also possible, however, to derive a ``hybrid'' expression, which combines aspects of both the mean-squared displacement and velocity autocorrelation approaches \cite{EN3193}.  This recently-derived alternative reads
\begin{equation}
\label{eq:difffhs}
D = \lim_{t\to \infty} \frac{1}{3}\avg{\vi(0) \cdot \Delta \ri(t)}.
\end{equation}
This expression yields an attractive physical interpretation: it states that the diffusion constant measures the extent to which a particle's initial velocity biases its eventual long-time displacement in the same direction.  For an isotropic medium (as is the case for liquids), one can rewrite the average in Equation \ref{eq:difffhs} as the integral of a joint probability distribution of initial velocity and final displacement.  Let $P(v_0,\Dx)$ be the probability that a particle has an initial $x$-component in velocity of $v_0$ and travels a distance $\Dx$ along the $x$-axis after a fixed time $\Dt$.  Thus $P$ is dependent on the particular time interval of interest $\Dt$, which will be implicit in our notation.  $P$ is independent, however, of the direction along which the the initial velocity and displacement are projected.  From a computational point of view, this means that each component of a velocity/displacement pair can be used to provide an independent estimate in the determination of $P$.  Using this joint probability distribution, the diffusion constant is then given by the following moment:
\begin{equation}
\label{eq:diffjointprob}
D = \lim_{\Dt\to \infty}  \int \int v_0 \cdot \Dx \cdot P(v_0,\Dx) d v_0 d\Dx .
\end{equation}
The function $P$ can be measured readily from a molecular dynamics trajectory; the positions and velocities are periodically saved, and after an interval $\Dt$ with respect to each such time origin, these values are used to update a two-dimensional histogram of initial velocity and final displacement \cite{EN3196}.  For theoretical progress, however, Equation \ref{eq:diffjointprob} is not particularly convenient in present form.  Instead, a more revealing expression arises when one integrates over $\Dx$.  Let $\Peq(v)$ be the equilibrium (Maxwell-Boltzmann) distribution of velocities along a single axis.  Taking the infinite-time limit, one obtains
\begin{equation}
\label{eq:diffcondprob}
D = \int v_0 \cdot \avg{\Dx|v_0} \cdot \Peq(v_0) d v_0
\end{equation}
where $\avg{\Dx|v_0}$ is the average distance travelled for the initial velocity $v_0$ in the limit $\Dt \to \infty$.  Clearly, symmetry demands that $\avg{\Dx|v_0}$ be an odd function of velocity.  The simplest suitable functionality is therefore a linear dependence, $\avg{\Dx|v_0} = C v_0$.  This is the form of the Stokes-Einstein approximation, for which $C=m/(6\pi \eta R)$ where $m$ is particle mass, $\eta$ is the shear viscosity, and $R$ is a particle hydrodynamic radius.

Naturally, the contributions of individual particles' motion to the diffusion constant in Equations \ref{eq:diffmsd}-\ref{eq:diffcondprob} fluctuate with some finite variance.  Normally in a hot liquid, differences in individual particle motions are not also reflected in their spatial positions.  In a dynamically heterogeneous environment, however, it is possible to choose a time interval $\Dt$ over which the more ``mobile'' and ``immobile'' particles cluster together.  The choice of $\Dt$ in this definition is important.  At extremely short times, the ballistic motion of particles is not spatially correlated, reflecting instead the Maxwell-Boltzmann distribution of velocities.  At very long times, on the other hand, particles must eventually lose memory of their original positions and velocities, and any spatial heterogeneity in the displacements is simply averaged out.  Thus the presence of dynamic heterogeneity implies the existence of an intermediate time scale, dependent on the temperature, which reveals clustering in terms of particle mobility.

The relevant quantities in studies of dynamic heterogeneity are the individual particle displacement vectors, $\Delta \ri (\Dt)$, and the associated scalars, $\Delta r_i (\Dt)$ .  In terms of these variables, ``mobile'' particles are those which have very large $\Delta r$, say greater than some cutoff value.  However, the precise and non-arbitrary selection of such a cutoff is often challenging.  One possibility is to choose the average displacement as a cutoff, but this can artificially enforce the mobile particles to be half the population of the entire system.  Another possibility is to examine particles whose $\Delta r$ lie at the extreme tails of the distribution, for example, the 5\% most mobile particles \cite{EN3113}.  This route has met considerable success in elucidating mechanisms of heterogeneous dynamics in simulations of supercooled liquids; however, it also contains an arbitrary parameter which affects the population of mobile particles.  In this work, we describe a technique for determining objectively the appropriate cutoff displacement from simulation calculations.  This approach utilizes the Cartesian-component version of the displacement vector, $\Dx(\Dt)$, and makes a direct connection with the self-diffusion formalism presented above in Equations \ref{eq:diffjointprob} and \ref{eq:diffcondprob}.

An interesting aspect of the intermediate-time particle displacement vectors of supercooled liquids is the existence of so-called non-Gaussian behavior \cite{EN3119}.  This refers to the probability distribution of displacements, $P(\Delta \ri)$, averaged over all particles and measured at the time $\Dt$.  Note that the scalar equivalent, $P(\Delta r)$, is the familiar van Hove correlation function. For purely random diffusion, $P(\Delta \ri)$ is Gaussian:
\begin{equation}
\label{eq:Gaussdist}
P(\Delta \ri) = \frac{1}{8(\pi D \Dt)^{3/2}}\exp\left(\frac{-|\Delta \ri|^2}{4 D \Dt}\right).
\end{equation}
At very short times, $\Delta \ri \approx \vi \Dt$ and thus $P$ is also Gaussian, due to the equilibrium distribution of velocities:
\begin{equation}
\label{eq:Gaussdist2}
P(\Delta \ri) = \left( \frac{m}{2 \pi k_B T (\Dt)^2} \right)^{3/2} \exp\left(\frac{-m|\Delta \ri|^2}{2 k_B T (\Dt)^2}\right)
\end{equation}
where $k_B$ is Boltzmann's constant and $T$ is the temperature.  It has been found that, for moderate to deeply supercooled liquids, the intermediate-time behavior of $P$ becomes substantially non-Gaussian, reflecting the effects of ``caged'' particles and the presence of dynamic heterogeneity.  The most frequently used indicator of non-Gaussian behavior is the parameter $\alpha_2$ \cite{EN3195}, which entails a ratio of the second and fourth moments of the displacement distribution:
\begin{equation}
\label{eq:alpha2}
\alpha_2(\Dt) = \frac{3 \avg{\Delta r(\Dt)^4}}{5 \avg{\Delta r(\Dt)^2}^2} - 1 .
\end{equation}
For a truly Gaussian distribution in $\Delta \ri$, $\alpha_2$ vanishes.  Thus, this non-Gaussian parameter is zero for $\Dt=0$, passes through a maximum at some intermediate time interval, and asymptotes back to zero as $\Dt \to \infty$.  To connect with the approach of Equations \ref{eq:diffjointprob} and \ref{eq:diffcondprob}, the expression for $\alpha_2$ is straightforwardly modified to incorporate the $x$-component distribution:
\begin{equation}
\label{eq:alpha2x}
\alpha_2(\Dt) = \frac{\avg{\Dx(\Dt)^4}}{5 \avg{\Dx(\Dt)^2}^2} - \frac{3}{5}  .
\end{equation}
Note that knowledge of the joint distribution of initial velocities and final displacements $P(v_0,\Dx)$, including its time dependence, immediately enables one to determine both the diffusion constant and the temporal evolution of the non-Gaussian parameter, since $P(\Dx)$ is simply the velocity integral of $P(v_0,\Dx)$.

\section{Computer simulations}
\label{sec:results}

We have performed molecular dynamics calculations for a well-studied binary mixture of Lennard-Jones particles (BMLJ) developed by Kob and Andersen \cite{EN3070}.  This model is a modification of a potential developed by Stillinger and Weber designed to mimic the behavior of a Ni$_{80}$P$_{20}$ metal-metalloid alloy \cite{EN773}, and it has been found to exhibit substantial caging, non-Gaussian behavior, and heterogeneous dynamics in the supercooled regime \cite{EN3069, EN3119, EN3120}.  The BMLJ system consists of two size- and energy- asymmetric types of particles, A and B, with equal mass.  All of the results reported here are for the more abundant A particles only, and all units are expressed in terms of the A-A interaction parameters and mass.  Our molecular dynamics calculations are performed in the NVE ensemble for a 250-particle system of 200 A and 50 B particles at total number density $\rho=1.2$, using the velocity Verlet algorithm with a time step of $\delta t = 0.003$.  Calculations are performed for two temperatures, $T=0.8$ and $0.5$, which correspond to moderately warm and supercooled states, respectively.  For comparison, a power-law fit shows that the diffusion constants in BMLJ extrapolate to zero around $T=0.435$ \cite{EN3069}.  Further details of the simulation methods can be found in Reference \cite{EN3196}.

\begin{figure}
\begin{center}
\includegraphics[width=\stdfigsize]{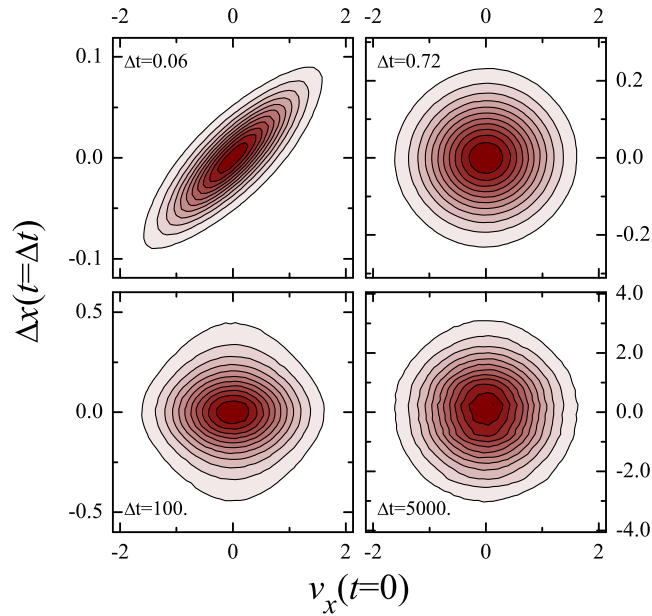}
\caption{\label{fig:fig1}
(color online) Calculated joint probability distributions $P(v_0,\Dx)$ for BMLJ at $T=0.5$, determined from molecular dynamics trajectories at four different time intervals.  $P(v_0,\Dx)$ gives the probability that a particle will have an initial velocity $v_0$ and travel a distance $\Dx$ after a time $\Dt$ later.  Each graph displays contour lines for this function, which are normalized relative to the maximum value.  For $\Dt=0.06$ the particle is still in the ballistic regime, with $\Dx$ strongly dependent on $v_0$.  Shortly thereafter, $P$ attains a Gaussian shape in both the $v_0$ and $\Dx$ directions.  At the intermediate time of $\Dt=100$, the contours develop a rhomboidal distortion, and eventually return to the Gaussian form at the longer time of $\Dt=5000$.}
\end{center}
\end{figure}

In Figure \ref{fig:fig1} we show contour depictions of the measured joint probability distributions $P(v_0,\Delta x)$ for the A particles in BMLJ at $T=0.5$.  This set of data is constructed by maintaining a two-dimensional histogram and using many time origins during the course of the simulation; around 10,000 time origins are employed in order to resolve the distributions to very high accuracy.  Figure \ref{fig:fig1} contains results for four time intervals spanning the ballistic to the long-time diffusive regime.  In the former, $\Dx$ is highly correlated with $v_0$ and the contour lines approach a skewed distribution along $\Dx = v_0 \Dt$.  Eventually this relationship becomes less obvious and the distribution appears axisymmetric; such decorrelation results from the inevitable interaction of a particle with its nearest neighbors when it continues along its initial velocity vector.  In fact, the distribution is nearly Gaussian, though it retains a very subtle skew such that the moment given by Equation \ref{eq:diffjointprob} remains non-zero, eventually approaching the diffusion constant.

Of particular note is the contour plot at $\Dt = 100$, which exhibits a pronounced rhomboidal distortion, i.e., a diamond-like shape.  This qualitative feature was previously identified as a characteristic of supercooled liquid dynamics, and was shown to persist even at times where particles had traveled nontrivial distances \cite{EN3196}.  At much longer times, random diffusive motion returns the contours to a Gaussian shape, as shown in the last plot of Figure \ref{fig:fig1} with $\Dt=5000$.  In the present work, it is quite striking that the diamond distortion emerges as an intermediate phenomenon between times at which the joint distribution is nearly Gaussian.  This may suggest a transition between two distinct mechanistic stages of diffusive motion.  Moreover, it is important to consider the dramatic separation of time scales over which the changes to the shape of the distribution occur (the ratio of times in Figure \ref{fig:fig1} being close to $1:10:10^3:10^5$).  In addition, the cross-moment of the distribution---which gives the diffusion constant---appears to converge to its infinite-time value by $\Dt = 100$ (results not shown; see Ref.~\cite{EN3196}).  This would seem to imply that the information relevant to the diffusion constant is already encoded in $P(v_0,\Dt)$ by the time of the distortion.  Although we do not show them here, we have also performed calculations of $P(v_0\Dx)$ for the higher temperature $T=0.8$.  In that case, an intermediate diamond-like shape is also found, although it is less pronounced and occurs at the earlier time of $\Dt = 1.44$.

\begin{figure}
\begin{center}
\includegraphics[width=\stdfigsize]{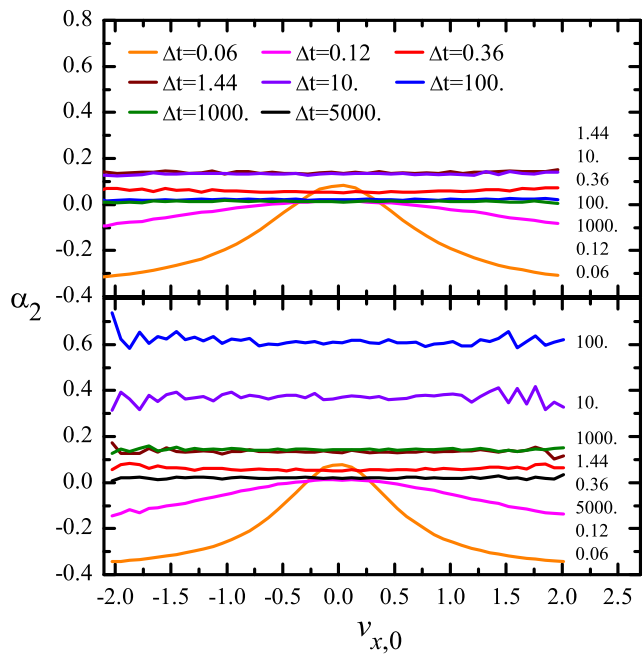}
\caption{(color online) Non-Gaussian parameter $\alpha_2$ as a function of initial particle velocity and time interval, for $T=0.8$ (top) and $T=0.5$ (bottom).  Except in the ballistic regime, a particle's initial velocity has little effect on its resulting non-Gaussian behavior.  The maxima in Gaussian deviation occur near $\Dt=1.44$ and $100$ for $T=0.8$ and $0.5$, respectively.}
\label{fig:fig2}
\end{center}
\end{figure}

Intuitively, one might anticipate that the diamond-shaped joint distribution at intermediate times in Figure \ref{fig:fig1} is related to a non-Gaussian distribution of particle displacements.  Thus a natural course of study is the examination of the non-Gaussian parameter $\alpha_2$ along each velocity ``slice'' in the distribution $P(v_0,\Delta x)$.  That is, one calculates the conditional distribution of particle displacements for each given initial velocity, and uses this subset of the complete displacement data to determine $\alpha_2$ using Equation \ref{eq:alpha2x}.  We have performed such calculations for a selection of time intervals $\Dt$ and for the two temperatures $T=0.8$ and $0.5$.  The results are displayed in Figure \ref{fig:fig2}.  Except at very short times during the ballistic regime, the calculations suggest that a particle's initial velocity has no effect on its subsequent non-Gaussian behavior---the values of $\alpha_2$ for $\Dt \ge 0.36$ are nearly uniform across the span of velocities.  This does not preclude the existence of small variations with initial velocity which may be obscured by statistical noise.  Recall, for instance, that a subtle correlation between initial velocity and displacement persists at all times (Equation \ref{eq:difffhs}).  The short-time non-Gaussian effects are readily rationalized: in this regime, the distribution of offsets for a particular initial velocity are sharply peaked around $\avg{\Dx|v_0}=v_0 \Dt$.  Thus the averages $\avg{\Dx^2}^2$ and $\avg{\Dx^4}$, when restricted to a given initial velocity, are nearly equal.  Substitution into Equation \ref{eq:alpha2x} yields the asymptotic value $\alpha_2(\Dt \to 0)=-2/5$.  For the special case of $v_0=0$, on the other hand, the short-time trajectory is determined by the particle accelerations, which are distributed according to the intermolecular forces.

Not surprisingly, the maximum in the non-Gaussian parameter is greater in amplitude and shifted to later times for the lower temperature of $T=0.5$.  This is in agreement with the earlier observations of Kob \etal \cite{EN3119}.  It is still interesting to note, however, that a noticeable maximum exists even at the relatively warm temperature of $T=0.8$.  Kob \etal proposed that the time corresponding to the $\alpha_2$ maximum is the interval relevant to dynamic heterogeneity.  We follow their approach and let this time be denoted $\Dt^*$, which takes values of $1.44$ and $100.0$ for $T=0.8$ and $0.5$, respectively.  Since we have only examined a small selection of $\Dt$ intervals, these values may be slightly offset from the location of the true maxima; nevertheless, they yield substantial non-Gaussian effects and are in reasonable agreement with earlier work \cite{EN3119}.  Although we don't present the results here, it is worthwhile noting that we find qualitatively similar behavior for the B particles, albeit with quantitative differences: the amplitudes of the B maxima in $\alpha_2$ are almost twice as great as for A particles and the corresponding times $\Dt^*$ occur slightly earlier.  This suggests that the B particles may be more sensitive indicators of spatial heterogeneity, which may be useful for future studies.

Returning to the A particles, we find that the time $\Dt^*$ coincides with the maximum diamond distortion in the joint probability plots of Figure \ref{fig:fig1}.  With this tentative empirical connection between the shape of $P(v_0,\Dx)$ and $\alpha_2$, the next step is to determine whether or not a non-Gaussian distribution of particle displacements alone is responsible for the diamond shape.  This appears to be the case.  As a simple exercise, we constructed \emph{a priori} a joint distribution in which the $\Dx$ axis consists of a sum of two Gaussian curves, perhaps the simplest non-Gaussian functionality.  Letting $G(z;\sigma)$ be a Gaussian distribution in $z$ centered at zero with standard deviation $\sigma$, our test function is given by $P(v_0,\Dx)=G(v_0;\sigma_v)[G(\Dx;\sigma_x) + G(\Dx;2\sigma_x)]/2$.  That is, the $\Dx$ component consists of a sum of two Gaussians whose standard deviation differs by a factor of two.  Rigorously speaking, this cannot be the true form of $P(v_0,\Dx)$, since it would imply a vanishing diffusion constant through Equation \ref{eq:diffjointprob}.  At longer times, however, the skew due to the diffusion constant is small, and thus the axisymmetric test form is not unreasonable.  Surprisingly, this very simple functional form completely reproduces the diamond shape in Figure \ref{fig:fig1}.  Thus, a non-Gaussian distribution of particle displacements appears to be the dominant factor in the rhomboidal distortion of the joint probability function.

\begin{figure}
\begin{center}
\includegraphics[width=\stdfigsize]{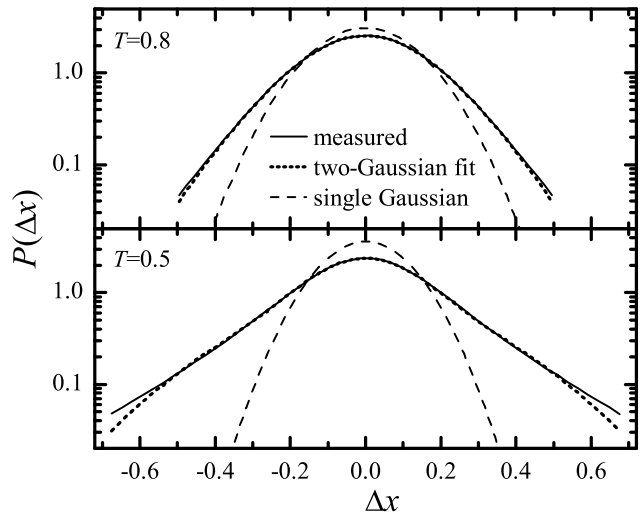}
\caption{The offset distributions $P(\Dx)$ and their two-Gaussian fits at the times of maximum non-Gaussian behavior.  The fits are of the form $P_\tx{fit}(\Dx) = f G(\Dx;\sigma_1) + (1-f) G(\Dx;\sigma_2)$ where $G$ is a Gaussian centered at zero, $\sigma$ is the standard deviation of the distribution, and $f$ weighs the relative contribution of the two Gaussians.  The single Gaussian curves correspond to $G(\Dx;\sigma_\tx{meas})$ where $\sigma_\tx{meas}$ is calculated from the measured probability distributions, $\sigma_\tx{meas} = \avg{\Dx^2}-\avg{\Dx}^2$.}
\label{fig:fig3}
\end{center}
\end{figure}

Perhaps more revealing, the two-Gaussian form works extremely well for describing the displacement probability at $\Dt^*$.   This is encouraging from a theoretical point of view, since two Gaussians might be associated with the physical existence of two spatially-separated, distinct diffusive environments in the liquid.  We explore this possibility in greater depth by performing a two-Gaussian fit of $P(\Dx)$ at all times $\Dt$ and for both temperatures studied.  We let the master function form be
\begin{equation}
\label{eq:twogaussfit}
P(\Dx) = f G(\Dx; \sigma_1) + (1-f) G(\Dx; \sigma_2)
\end{equation}
where $0 \le f \le 1$ is the relative weight of the first Gaussian.  The fit is performed by minimizing the total squared difference between the measured and fit functions, $\sum_{\Dx} [P_\tx{fit}(\Dx)-P(\Dx)]^2$, with respect to the fit parameters $\sigma_1$, $\sigma_2$, and $f$.  We have verified that different initial guesses for these parameters do not affect their final values after the minimization procedure.  Figure \ref{fig:fig3} shows the results of the fits at $\Dt^*$ for the two temperatures studied.  For comparison, a single Gaussian approximation is also shown, using the measured standard deviation of particle displacements.  The two-Gaussian form reproduces the actual $P(\Dx)$ distributions very well, with the exception of extreme values of $\Dx$ whose weight it underestimates.  If one assumes that diffusive motion occurs largely in discrete ``hopping'' events, these deviations might correspond to those rare particles that hop twice in the time interval.  Thus the true distribution might entail a third and higher-order Gaussians to describe multiple hops.  In this sense, the particle displacement probability at intermediate times might be described as an infinite sum of Gaussians of decreasing magnitude, each corresponding to particles that hop $n$ times.  This will be the subject of future work.

\begin{figure}
\begin{center}
\includegraphics[width=\stdfigsize]{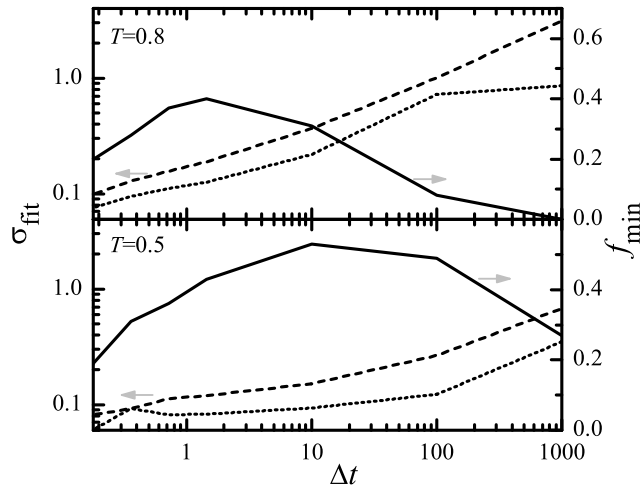}
\caption{Fit parameters for the two-Gaussian regression.  The left axis gives the values of the standard deviations of the two distributions, $\sigma_1$ and $\sigma_2$, while the right axis gives the relative weight $f$ of the Gaussian with the smaller $\sigma$. }
\label{fig:fig4}
\end{center}
\end{figure}

The parameters resulting from the fitting procedure are shown in Figure \ref{fig:fig4}.  In the ballistic and long-time regimes at both temperatures, the measured $P(\Dx)$ are almost perfectly Gaussian, and thus the two-Gaussian fit is ill-defined.  This means that either the weight $f$ approaches 0 or 1, or the two standard deviations become nearly equal.  Therefore, Figure \ref{fig:fig4} contains the fitted parameters only for the intermediate times of $\Dt=0.18$ to $1000$.  As one would expect, the maximum in $f$, which weighs the relative contribution of the Gaussians, occurs close to the time of maximum non-Gaussian behavior (see Figure \ref{fig:fig2} for comparison).  Similar to the results for $\alpha_2$, the maximum in $f$ at the lower temperature is greater in magnitude and occurs at later times.  Near the $f$ maximum, the ratio of the two $\sigma$ parameters is roughly $1.5$ for $T=0.8$ and $2.0$ for $T=0.5$.  This ratio remains in the range $1.3-2.2$ for all times such that $f$ is greater than $0.3$.  This range encompasses the value of $\sqrt{2}$ which would be obtained if $\sigma_1$ and $\sigma_2$ corresponded to purely random single and double ``hops'' of the same length.  Also note that in three dimensions, $\avg{\Delta r^2} = 3 \avg{\Dx^2}$ and so the equivalent $\sigma$ parameters for diffusion along all three axes entails an additional factor of $\sqrt{3}$.

The question immediately arises as to whether the contributions to each of the two Gaussians are spatially correlated.  To investigate this, we define a critical value of $\Dr^*$ corresponding to the average of the standard deviation of the two distributions: $\Dr^* = \sqrt{3} (\sigma_1+\sigma_2)/2$.  This is not the only possible definition of $\Dr^*$ in terms of the two Gaussians; for example, one might choose the point at which $f G(\Dr; \sqrt{3} \sigma_1) = (1-f) G(\Dr; \sqrt{3} \sigma_2)$.  Another possibility is to let mobile particles be those with $\Dr$ values greater than the larger $\sigma$, and immobile particles those with $\Dr$ less than the smaller $\sigma$.  We choose the average for its simplicity.  Subsequently, we define particles which travel farther than $\Dr^*$ during the time period $\Dt$ as mobile and the rest as immobile.  This is similar to the analysis of Kob \etal \cite{EN3119}, who defined the critical distance in terms of deviations from a single-Gaussian fit.  Our approach has the advantage that the length scales $\sigma_1$ and $\sigma_2$ are intrinsic to the diffusive mechanisms in the liquid.  It is also important to note that we do not require that mobile or immobile particles constitute a fixed fraction of the total number, which enables one to examine how their number fluctuates with time.

\begin{figure}
\begin{center}
\includegraphics[width=\widefigsize]{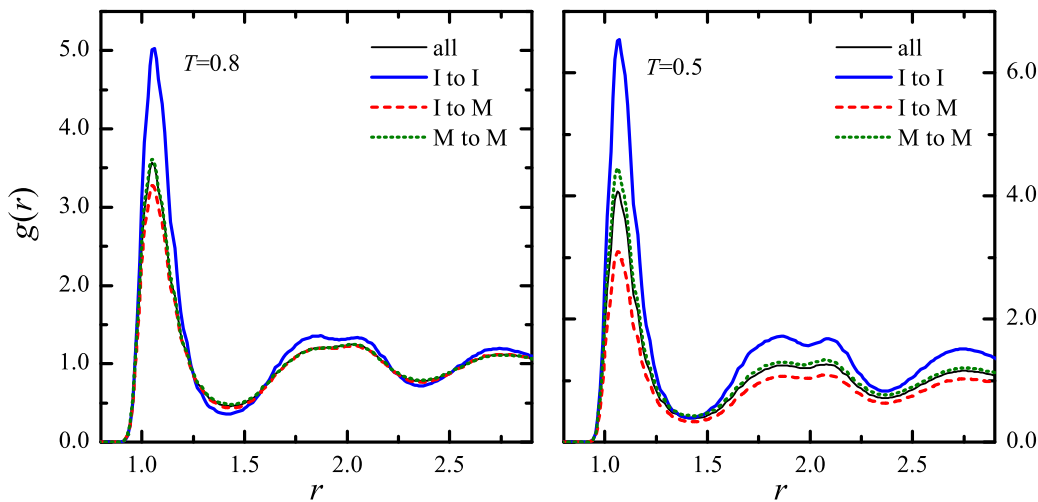}
\caption{(color online) Pair correlation functions for mobile and immobile A particles in BMLJ.  Mobile particles are those whose displacement is greater than $\sqrt{3}(\sigma_1+\sigma_2)/2$ in the two-Gaussian fit; immobile particles are the remainder.}
\label{fig:fig5}
\end{center}
\end{figure}

Figure \ref{fig:fig5} contains the pair correlation functions $g(r)$ for mobile and immobile A particles at the time of maximum non-Gaussian behavior in BMLJ.  These plots are constructed using a large number of time origins, and thus represent the average populations of each type of particle.  At both temperatures, the immobile-to-mobile correlations are the weakest, suggesting a spatial separation according to mobility. Particularly striking is the strong correlation between immobile particles, which is greater than its mobile-to-mobile counterpart and which grows as the temperature is lowered.  In fact, at both temperatures the mobile-mobile pair correlation function remains close to that of the average A-A interactions.  Intuitively, one might expect the immobile correlations to be stronger, since at low temperature these particles likely vibrate around fixed equilibrium positions during the time interval, whereas mobile particles are, by definition, more transient in their positions.  Another interesting feature borne out in Figure \ref{fig:fig5} is the broad shoulder present in the second peak of the pair correlation functions, which appears exaggerated in the case of the immobile-immobile relationships.  This feature was identified as a structural precursor to freezing in work by Truskett \etal \cite{EN675}; a tentative connection could be that immobile regions in the liquid are ``well-packed'' with respect to each other, lowering their mobility.  Still, Ref.~\cite{EN675} investigated single-component systems, and thus the generality of their conclusions regarding the double-peaked structure remains to be verified in mixtures such as BMLJ.

\section{Discussion and conclusions}
\label{sec:conclusion}

The wealth of dynamic phenomena that occur in supercooled liquids would seem to hinder a complete understanding of their behavior, and indeed the development of a general theory of these substances remains challenging.  However, connections between various dynamic anomalies are being established, such as those which are the focus of the present work, non-Gaussian diffusion and dynamic heterogeneity.  In some sense, much progress in this field has occurred by maintaining a diversity of theoretical perspectives and simulation techniques, although a satisfactory comprehensive viewpoint is still lacking.  Our contribution in the present work has been to provide a new approach to studying dynamic heterogeneity, which is based on a rigorous theory for the diffusion constant and which is readily measured in molecular dynamics simulations.

We have investigated the familiar binary mixture of Lennard-Jones particles as a model supercooled liquid.  The main focus of our efforts has been the probability distribution of particle displacements in a given time window, including their dependence on initial velocity.  We have shown that the joint distribution of initial velocity and displacements attains a characteristic diamond shape in supercooled liquids, at the time of maximum non-Gaussian behavior.  Furthermore, these results indicate that a particle's initial velocity is not correlated with the degree of non-Gaussian behavior it exhibits.  Motivated by the observed diamond shape, we have found that the distribution of particle displacements at intermediate times can be almost completely reproduced by the sum of two Gaussians with standard deviations that differ in value by a factor of 1.3--2.2.  Moreover, these two length scales provide an objective mechanism for identifying ``mobile'' and ``immobile'' particles.  An analysis of spatial correlations in the liquid indicates that the two Gaussians correspond to mobile and immobile regions.

The two-Gaussian fit allows the identification of length scales of mobile and immobile particles, in terms of their displacement.  This permits the development of mobility criteria which do not entail an arbitrary compositional constraint, such as assigning as mobile the 5\% of particles with the greatest displacement.  Without the constraint, it is possible to examine the temporal evolution of the number of mobile particles, for instance, by tracking a dynamic trajectory relative to itself at an earlier time.  The instantaneous mobile population is also in a sense heterogeneous, in that it experiences large variations as time progresses.  A useful future study might characterize the distribution in time of local mobile populations, that is the number of mobile particles in small subvolumes of the system.

The recent results of Widmer-Cooper \etal \cite{EN3191} also appear promising as a method for identifying dynamically distinct regions.  These authors perform numerous molecular dynamics trajectories using a common initial configuration, but with different initial velocities.  They demonstrate that, when averaged over the possible trajectories consistent with the system's temperature, certain regions of particles are more likely to diffuse than others.  The implication of their work is the apparent existence of structural (static) distinctions which yield heterogeneous dynamic behavior, and it would be interesting to investigate the correspondence between types of particles identified by their method and ours.

The two-Gaussian perspective provides a useful physical interpretation.  It suggests that the transient regions of mobility and immobility that emerge in the supercooled liquid can be described by distinct diffusion constants.  If the dynamics are described as ``hopping'' events, the Gaussian with the larger length scale might then correspond to clusters or strings of particles which are able to make a collective hop.  Over long time scales, the regions themselves change---in terms of their physical location in the system and their constituent particles---and distinctions among particles become averaged out.  In this interpretation, therefore, the intermediate time scale at which two Gaussians exist must be related to the rates at which mobile and immobile regions interchange, since at longer times individual particles experience multiple diffusive environments and thus possess an averaged displacement.

Our present analysis does not yet explain the manner in which regions of varying mobility emerge in the liquid.  Future investigations will attempt to understand the origin of the multiple diffusive length scales by tracking their development as the hot liquid is cooled.  More generally, the connection between such dynamically-defined regions and basic features of interparticle forces remains to be established.  Identifying a correlation between multiple length scales and the statistical properties of a system's multidimensional potential energy landscape, for instance, would represent a major theoretical advance.  In particular, the concept of energy landscape ``metabasins'' \cite{EN3078} may be a promising route for extending our approach.

\ack
P.G.D. gratefully acknowledges financial support by the U.S. Department of Energy, Division of Chemical Sciences, Geosciences, and Biosciences, Office of Basic Energy Sciences, Grant No. DE-FG02-87ER13714.  M.S.S gratefully acknowledges the support of the Fannie and John Hertz Foundation.

\end{document}